\documentclass[runningheads]{llncs}
\usepackage[dvipsnames]{xcolor}
\usepackage{graphicx}
\usepackage{comment}
\begin{document}
\title{FocalErrorNet: Uncertainty-aware focal modulation network for inter-modal registration error estimation in ultrasound-guided neurosurgery}
\titlerunning{FocalErrorNet for inter-modal registration error estimation}

\author{Soorena Salari 
% index{Salari, Soorena}
\inst{1} \and
Amirhossein Rasoulian 
% index{Rasoulian, Amirhossein}
\inst{1}\and
Hassan Rivaz
% index{Rivaz, Hassan}
\inst{2}\and
Yiming Xiao
% index{Xiao, Yiming}
\inst{1}}
\authorrunning{S. Salari et al.}
\institute{Department of Computer Science and Software Engineering, Concordia University, Montreal, Canada\\
\email{\{soorena.salari,ah.rasoulian,yiming.xiao\}@concordia.ca}
\and
Department of Electrical and Computer Engineering, Concordia University, Montreal, Canada\\
\email{hassan.rivaz@concordia.ca}}

\maketitle              % typeset the header of the contribution
\begin{abstract}
In brain tumor resection, accurate removal of cancerous tissues while preserving eloquent regions is crucial to the safety and outcomes of the treatment. However, intra-operative tissue deformation (called brain shift) can move the surgical target and render the pre-surgical plan invalid. Intra-operative ultrasound (iUS) has been adopted to provide real-time images to track brain shift, and inter-modal (i.e., MRI-iUS) registration is often required to update the pre-surgical plan. Quality control for the registration results during surgery is important to avoid adverse outcomes, but manual verification faces great challenges due to difficult 3D visualization and the low contrast of iUS. Automatic algorithms are urgently needed to address this issue, but the problem was rarely attempted. Therefore, we propose a novel deep learning technique based on 3D focal modulation in conjunction with uncertainty estimation to accurately assess MRI-iUS registration errors for brain tumor surgery. Developed and validated with the public RESECT clinical database, the resulting algorithm can achieve an estimation error of 0.59±0.57 mm.

\end{abstract}
\section{Introduction}

Resection of early-stage brain tumors can greatly reduce the mortality rate of patients. During the surgery, brain tissue deformation (called brain shift) can occur due to various causes, such as gravity, drug administration, and pressure change after craniotomy. While modern magnetic resonance imaging (MRI) techniques can provide rich anatomical and physiological information with various contrasts (e.g., fMRI) for more elaborate pre-surgical planning, intra-operative MRI that can track brain shift requires a complex setup and is costly. In contrast, intra-operative ultrasound (iUS) has gained popularity for real-time imaging during surgery to monitor tissue deformation and surgical tools because of its lower cost, portability, and flexibility \cite{rivaz2015near}. Accurate and robust MRI-iUS registration techniques \cite{xiao2019evaluation} can greatly enhance the value of iUS for updating pre-surgical plans and guiding the interpretation of iUS, which has an unintuitive contrast and non-standard orientations. This can greatly enhance the safety and outcomes of the surgical procedure by allowing maximum brain tumor removal while avoiding eloquent regions \cite{marko2014extent}. However, as the true underlying tissue deformation is unknown due to the 3D nature of the surgical data and the time constraint, real-time manual inspection of MRI-iUS registration results is challenging and error-prone, especially for precision-sensitive neurosurgery. Therefore, algorithms that can detect and quantify unreliable inter-modal medical image registration results are highly beneficial. 
\\
\indent Recently, automatic quality assessment for medical image registration has attracted increasing attention \cite{bierbrier2022estimating} from the domains of big medical data analysis and surgical interventions. With high efficiency, machine, and deep learning techniques have been proposed to allow automatic grading and dense estimation of medical image registration errors. Early endeavors on this topic primarily relied on hand-crafted features, including information theory-based metrics \cite{shams2017assessment,saygili2018local,sokooti2019quantitative,sokooti2016accuracy,eppenhof2018error,sokooti2021hierarchical}. More recently, deep learning (DL) techniques that learn task-specific features have also been adopted in automatic evaluation of medical image registration, with a primary focus on intra-contrast/modal applications, including CT \cite{eppenhof2018error,sokooti2021hierarchical} and MRI \cite{fonov2022darq}. Unfortunately, so far, error grading and estimation in inter-contrast/modal registration have rarely been explored, despite the particular demand in surgical applications. In this direction, Bierbrier \textit{et al.} \cite{bierbrier2023towards} made the first attempt using simulated iUS from MRI to train 3D convolutional neural networks (CNNs) to perform dense error regression for MRI-iUS registration in brain tumor resection. Although their algorithm performed well in simulated cases, the results on real clinical scans still required improvements. In this paper, we propose a novel 3D CNN to perform patch-wise error estimation for MRI-iUS registration in neurosurgery, by using focal modulation \cite{yang2022focal}, a recent alternative DL technique to self-attention \cite{AttentionAllYouNeed} for encoding contextual information, and uncertainty estimation. We call our method FocalErrorNet, which has three main novelties. \textbf{First}, we adapted the focal modulation network \cite{yang2022focal} from 2D to 3D and employed the technique in registration error assessment for the first time. \textbf{Second}, we incorporated uncertainty estimation using Monte Carlo (MC) dropouts \cite{gal2016dropout} to offer assurance for error regression. \textbf{Lastly}, we developed and thoroughly evaluated our technique against a recent baseline model \cite{bierbrier2023towards} using real clinical data and showed excellent results. 

\section{Methods and Materials}
\subsection{Dataset and preprocessing}
\begin{figure*}[h]
\centering
\includegraphics[scale=0.4]{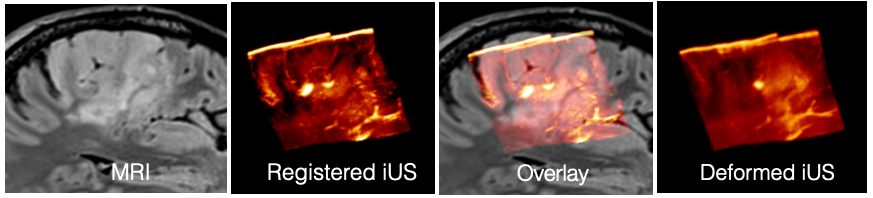}
\caption{Left to right: demonstration of sample pre-operative MRI, perfectly registered, and deformed iUS with a mean registration error of 1.4 mm.}
\label{SampleMRDef}
\end{figure*}

For methodological development and assessment, we used the RESECT (REtroSpective Evaluation of Cerebral Tumors) dataset \cite{xiao2017re}, which has pre-operative MRI, and iUS scans at different surgical stages from 23 subjects who underwent low-grade glioma resection surgeries. As it is still challenging to model iUS scans with tissue resection, we took 22 cases with T2FLAIR MRI that better depicts tumor boundaries and iUS acquired before resection. An example of an MRI-iUS pair from a patient is shown in Fig. \ref{SampleMRDef}. We hypothesized that directly leveraging clinical iUS could help learn more realistic image features with potentially better outcomes in clinical applications than with simulated contrasts \cite{eppenhof2018error,bierbrier2023towards}. However, since the true brain shift model is impossible to obtain, we followed the strategy of creating silver ground truths for image alignment \cite{eppenhof2018error,bierbrier2023towards}, upon which simulated misalignment is augmented in the iUS to build and test our DL model. To create the silver registration ground truths, we used the homologous landmarks between MRI and iUS in the RESECT dataset to perform landmark-based 3D B-Spline nonlinear registration to register iUS to the corresponding MRI for all 22 cases. To tackle the limited field of view (FOV) in iUS, we cropped the T2FLAIR MRI to the same FOV of the iUS, which was resampled to a $0.5 \times 0.5 \times 0.5$ $\mathrm{mm}^3$ resolution. 

\begin{comment}
\begin{figure*}[h]
\centering
\includegraphics[scale=0.6]{Images/Resect_Explanation.png}
\caption{Illustration of sample MRI-iUS of the RESECT dataset.}
\label{SampleMRUS}
\end{figure*}
\end{comment}

To perform spatial misalignment augmentation, we continued to leverage 3D B-Spline transformation, similar to earlier reports on the same topic \cite{lotfi2013improving,sokooti2021hierarchical,bierbrier2023towards}. In short, B-Spline transformation can be modeled by a grid of regularly spaced control points and the associated parameters to allow various levels of nonlinear deformation. While the spacing of the control points determines the levels of details in local deformation fields, the displacement parameters control the magnitude of the deformation. To ensure that simulated registration errors are of different varieties and sizes, we randomly selected the number of control points and the associated displacements (in each 3D axis) with a maximum of 20 points and 30 mm, respectively. Note that the control point grid is isotropic, and the density is arbitrarily determined per deformation in our case. Each co-registered iUS scan was deformed ten times. After misalignment augmentation on the previously co-registered iUS, matching pairs of 3D image patches of size $33 \times 33\times 33$ voxels were taken from both the iUS volume and the corresponding MRI.  As iUS has limited FOV and may contain no anatomical features, to ensure that the patches we extracted contain useful information (e.g. to avoid the dark background) in iUS, we focused on acquiring patches centered around the anatomical landmark locations available through the RESECT database. Since B-spline transformation offers a displacement vector at each voxel of the iUS volume, we directly considered the norm of the vector as the simulated registration error at the associated voxel. In our design, we determined the registration error of the image patch pair as the mean of all voxel-wise errors within the iUS patch. Finally, the image patch pairs, along with corresponding registration errors were then fed to the proposed DL algorithm for training and validation.

\subsection{Network architecture}
\begin{figure*}[h]
%\centering
\includegraphics[scale=0.28]{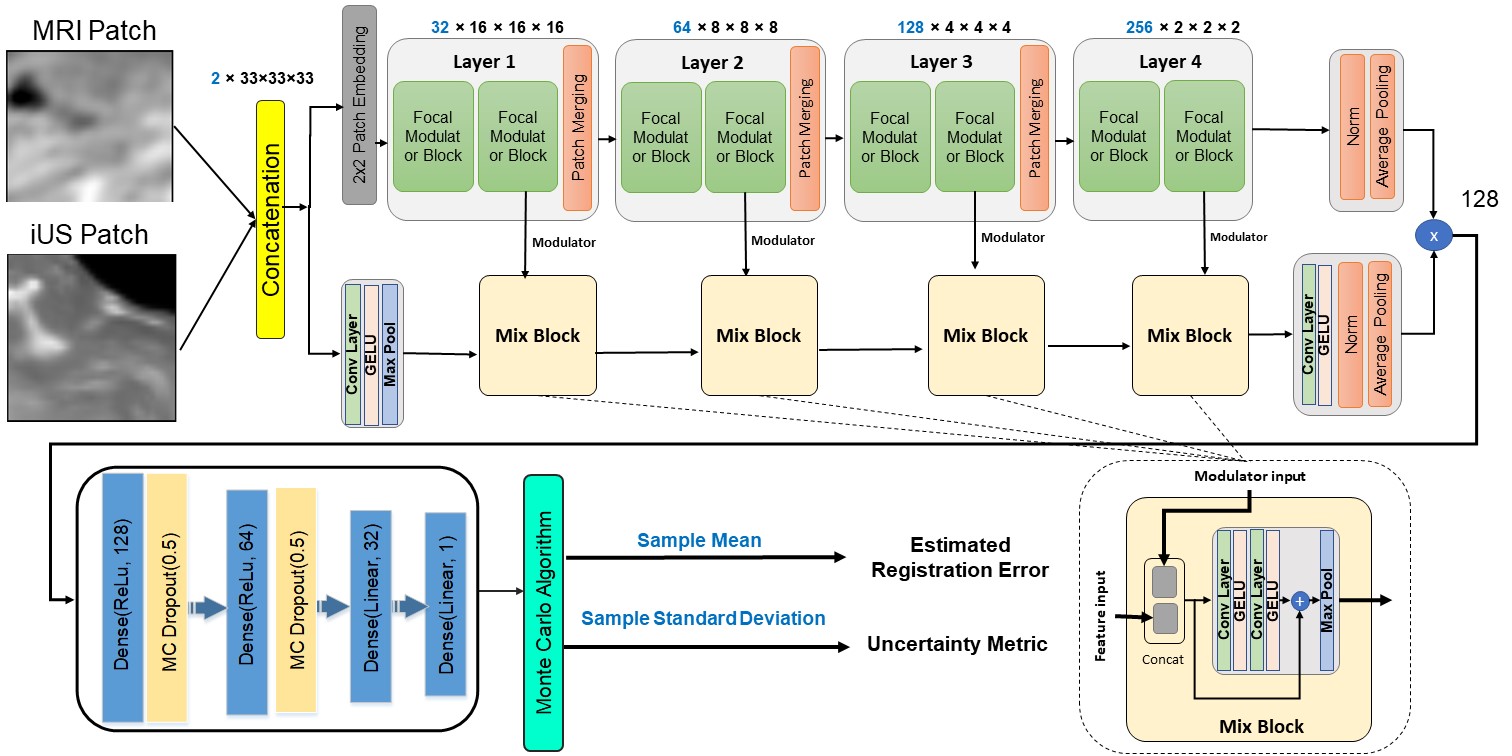}
\caption{The proposed FocalErrorNet for registration error and uncertainty estimation.}
\label{FocalErrorNet}
\end{figure*}

We proposed a novel 3D neural network, named FocalErrorNet, based on the recent focal modulation networks \cite{yang2022focal} that was originally proposed for 2D vision tasks to estimate the registration error between MRI and iUS patches. With a similar goal as the Vision Transformer (ViT), the focal modulation network was designed to model contextual information in images. It incorporates three main elements to achieve the goal: 1) focal contextualization that comprises a stack of depth-wise convolutional layers to account for long- to short-range dependencies, 2) gated aggregation to collect contexts into a modulator for individual query tokens, and 3) element-wise affine transformation to inject the modulator into the query. In the architecture of FocalErrorNet (see Fig. \ref{FocalErrorNet}), all layers contain two focal modulator blocks, where two depth-wise convolutional layers focally extract contexts around each voxel, selectively aggregate and inject them into the query, and pass the information to the next block.  We designed the FocalErrorNet as a ResNet-like variant of the focal modulation network to better encode relevant features across the input image and ensure a better gradient flow. Finally, the information from the backbone was propagated to a multi-layer perceptron (MLP) to regress registration errors, and two MC dropout layers were added to the MLP to allow uncertainty quantification for the results.

\begin{figure*}[h]
%\centering
\includegraphics[scale=0.2]{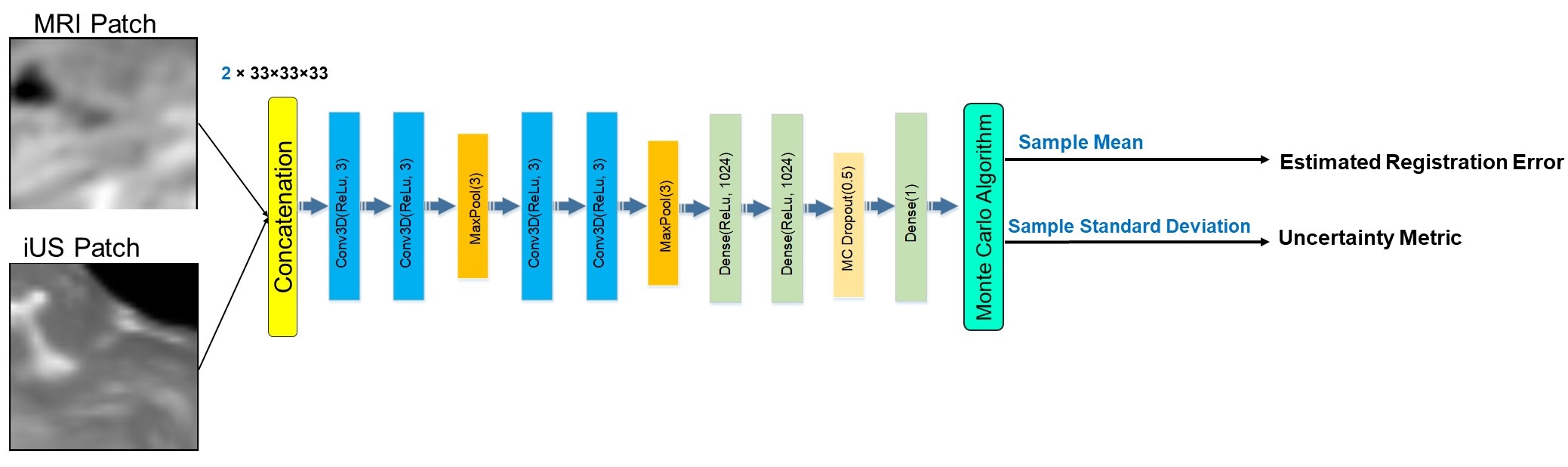}
\caption{The baseline 3D CNN \cite{bierbrier2023towards,eppenhof2018error} with the added MC dropout layer.}
\label{Model_3DCNN}
\end{figure*}

\subsection{Uncertainty quantification}
For registration error regression in surgical applications, knowledge regarding the reliability of the automated results is instrumental for the safety and well-being of the patients. Uncertainty estimation has gained popularity in probing the trustworthiness and credence of DL algorithms. Although the concept has been widely applied in image segmentation and classification, it has not been employed for registration error estimation, especially in the case of multi-modal situations, such as MRI-iUS alignment. Therefore, we incorporated uncertainty estimation in our proposed FocalErrorNet. For each MRI-iUS patch pair, 200 regression samples were collected by random sampling from MC dropouts \cite{gal2016dropout} at test time. While the final patch registration error was obtained as the mean of all the samples, the sample standard deviation was used as the uncertainty metric.   

\begin{table}[h]
%\hspace{-30cm}
\centering

\caption{Accuracy comparison of different models for registration error estimation.}
\begin{tabular}{ccc p{3cm} p{3.9cm} p{2.5cm} p{1.1cm} p{1.5cm} p{2.5cm}}
\hline

Model & Absolute error (mm)  & Correlation \\

%Mean absolute error & Mean squared error\\
\hline
\hline

3D CNN \cite{eppenhof2018error,bierbrier2023towards} &  1.69$\pm$1.37  & 0.61  \\
\hline

%\textbf{Swin-V2} &  \textbf{4.92$\pm$0.4/4.18$\pm$0.3} \\%& \textbf{54.60/45.41}\\
\textbf{FocalErrorNet} & \textbf{0.59$\pm$0.57}  & \textbf{0.82}   \\

%6.03 &  40.21%\\
%\emph{Proposed ViT model} & \textbf{99.4}\% \\
\hline
\end{tabular}
\label{Comparison}
\end{table}

\subsection{Experimental setup and implementation}
From the transformation augmentation, we acquired 3380 samples of MRI-iUS pairs. For our experiments, we arbitrarily split the subjects into training, validation, and test sets with the proportion of 60\%, 20\%, and 20\%, respectively. To prevent information leakage, we ensured that each patient was included in only one of the split sets. For model training, we adopted the Adam optimization with a learning rate of $5 \times 10^{-5}$ and a batch size of 64. For the loss function, we used mean squared error (MSE) to minimize the difference between the predicted MRI-iUS registration error and the ground truths. Furthermore, in addition to the transformation augmentation, we also included additional data augmentation, including random noise addition and random image flipping on training sets to mitigate overfitting and increase the model's generalizability. To assess our proposed FocalErrorNet, we compared it against a 3D CNN \cite{eppenhof2018error,bierbrier2023towards} (see Fig. \ref{Model_3DCNN}) that was employed for medical image registration error regression. The two DL models were trained with the same dataset and procedure, and their prediction accuracies, measured as the absolute error between the predicted and ground truths mis-registration on the test set were compared with two-sided paired-samples t-tests to confirm the superiority of the proposed method, in addition to correlations between their estimated and ground truth errors. To validate the proposed uncertainty estimation method, we calculated the correlation between the uncertainty measure and absolute error of FocalErrorNet, and the correlation between the uncertainty and mutual information between MRI and iUS, which is often used to measure the information overlap in multi-modal registration. Finally, to test the robustness of the FocalErrorNet, we acquired additional MRI-iUS patch pairs from the test subjects, by introducing random linear shifts (the max displacement from landmark locations is 10 voxels) from the selected locations in the original set, and evaluated the DL model performance.

\begin{figure*}[h]
\centering
\includegraphics[scale=0.36]{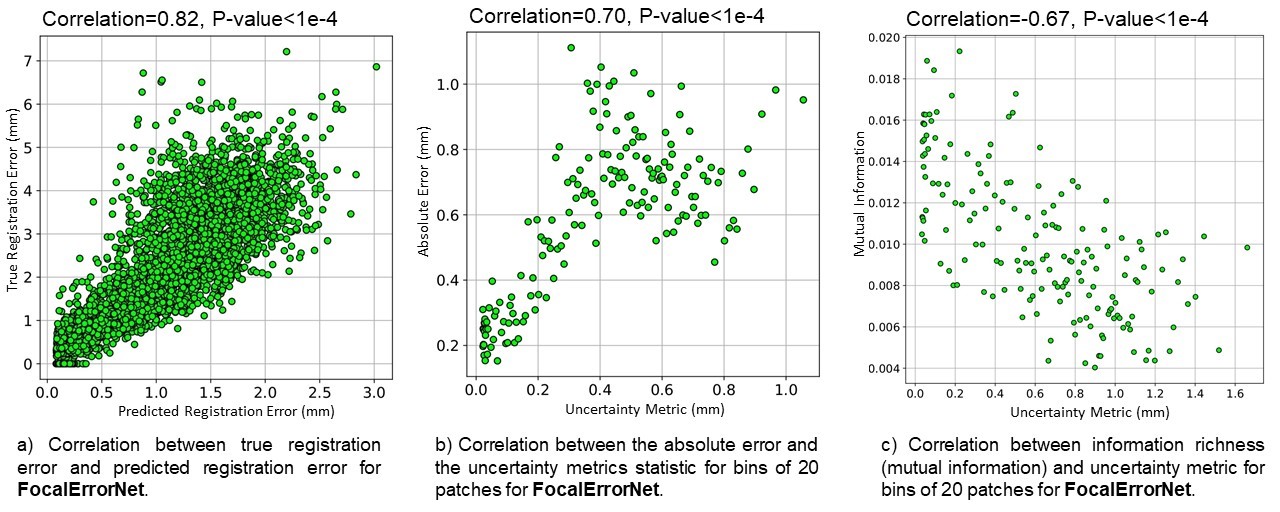}
\caption{Left to right: scatter plots and correlations between true registration error vs. predicted registration error, uncertainty metric vs. absolute error, and mutual information vs. uncertainty metric for the proposed FocalErrorNet.}
\label{FocalErrorNet_Scatter_Plots}
\end{figure*}
\begin{figure*}[h]
\centering
\includegraphics[scale=0.35]{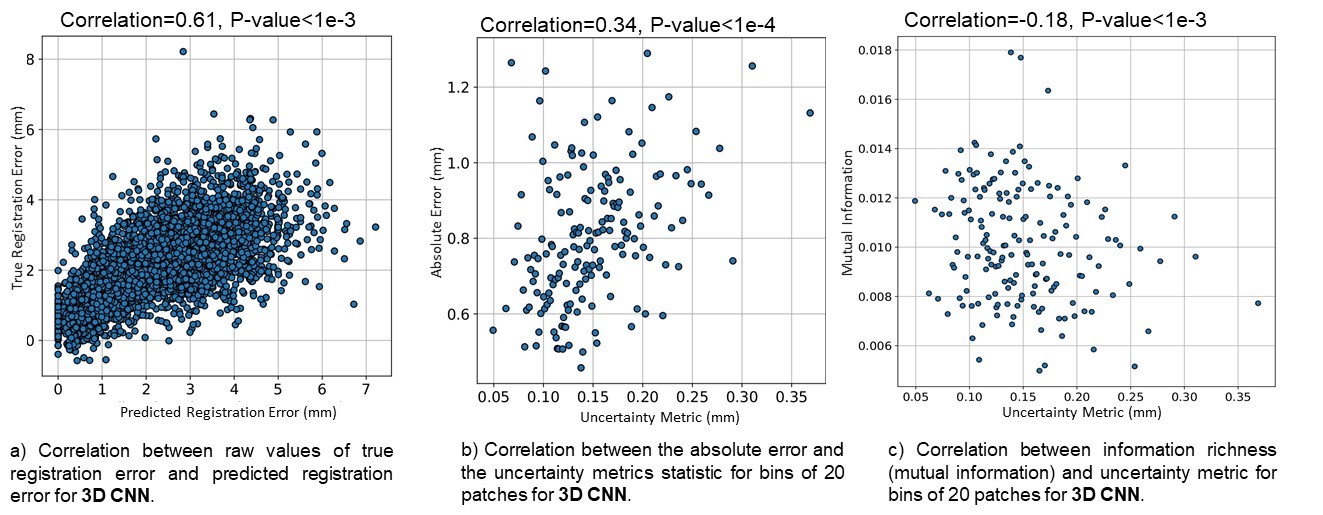}
\caption{Left to right: scatter plots and correlations between true registration error vs. predicted registration error, uncertainty metric vs. absolute error, and mutual information vs. uncertainty metric for the 3D CNN \cite{eppenhof2018error,bierbrier2023towards}.}
\label{3D_CNN_Scatter_Plots}
\end{figure*}

\section{Results}
\subsection{Error regression accuracy}
The accuracy comparison between the proposed FocalErrorNet and the baseline 3D CNN \cite{eppenhof2018error,bierbrier2023towards} is shown in Table \ref{Comparison}. Across all samples in the testing data, we achieved an accuracy of 0.59±0.57 mm, while the counterpart obtained a prediction error of 1.69±1.37 mm. With the t-test, our FocalErrorNet outperformed the 3D CNN \cite{eppenhof2018error,bierbrier2023towards} (p$<$1e-4). In addition, the correlations between the predicted and ground truths errors are 0.82 (p$<$1e-4) and 0.61 (p$<$1e-3) for FocalErrorNet and 3D CNN, respectively, further confirming the advantage of the proposed technique. To allow a qualitative comparison, scatter plots for predicted vs. ground truth errors of the two models are depicted in Fig. \ref{FocalErrorNet_Scatter_Plots}a and \ref{3D_CNN_Scatter_Plots}a. At larger error levels, it is evident that the point clouds exhibit a wider  shape.

\subsection{Validation of the uncertainty evaluation}
\label{Validating_UNC_Metric}

 We obtained correlations of 0.70 (p$<$1e-4) and 0.34 (p$<$1e-4) between estimated uncertainty and prediction error for FocalErrorNet and the baseline 3D CNN, respectively. Additionally, the uncertainty vs. mutual information uncertainties was assessed at -0.67 (p$<$1e-4) for our proposed method and -0.18 (p$<$1e-3) for the baseline. To allow better visual comparisons, the associated scatter plots are illustrated in Fig. \ref{FocalErrorNet_Scatter_Plots} and \ref{3D_CNN_Scatter_Plots}. These metrics proved the validity of our uncertainty measure and further confirmed the performance of FocalErrorNet. Note the scatter plots for uncertainty measure validation were performed using value binning (with 20 values per bin) for each axis to better reveal the trends of the metrics.

\subsection{Robustness of the proposed model}

To examine the performance of our proposed method for image regions that contain fewer potent anatomical features, we acquired additional image pairs from test subjects, according to Section 2.4. With the new test set, the prediction errors for our method and the baseline model were 1.28±0.99 mm and 2.49±1.87mm, respectively. Furthermore, the correlations between estimated and true error were calculated at 0.41 for FocalErrorNet and 0.20 for the baseline. These results supported the benefits of focal modulation in registration error estimation. In this test, patches can contain large areas of zeros (image content out of the scanning FOV of the iUS). The main reason for the observed performance decline is due to the reduction in sufficient image features in iUS. However, despite these challenges, we saw an acceptable outcome from FocalErrorNet (absolute error=1.28mm or $\sim$1 voxel in clinical MRIs).

\section{Discussion}
In image-guided interventions, there is an urgent need for automatic assessment of image registration quality. Multi-modal registration quality evaluation poses major challenges due to three main factors. First, dissimilar contrasts between images require more elaborate strategies to derive relevant features for error assessment. Second, unlike segmentation or classification, the ground truths of registration errors are difficult to obtain. Finally, compared with classification, regression tasks tend to be more error-prone for deep learning algorithms. To tackle these challenges, we employed 3D focal modulation with depth-wise convolution to encode contextual information for the image pair. Compared with the ViT and its variants, focal modulation allows a more lightweight setup, which could be desirable for 3D data. Although we admit that residual errors still remain after landmark-based B-Spline nonlinear alignment, this approach has been adopted in different prior studies, considering the residual landmark registration error is fairly low (mTRE of 0.0008±0.0010mm). Although simulated ultrasound has been used to provide a perfect alignment with MRIs, the fidelity of the simulated results is still suboptimal, and this may explain the under-performance of the previous technique in real clinical data \cite{bierbrier2023towards}. To ensure the performance of our FocalErrorNet, we opted to regress the mean registration error of image patches than simplistic error grades or voxel-wise error maps. We believe that this design choice offers a more stable performance, which is supported by our validation. We adopted uncertainty estimation in inter-modal registration error assessment for the first time. While other techniques exist to provide model uncertainty \cite{abdar2021review}, MC dropout is more flexible for various DL models. Furthermore, the use of standard deviation as an uncertainty measurement maintains the same unit as the regressed errors, thus making the interpretation more intuitive. From quantitative and qualitative evaluations using correlation coefficients and scatter plots to assess the association of uncertainty measures with the prediction errors and image entropy, we confirmed the validity of the proposed uncertainty estimation approach. For our FocalErrorNet, we achieved a prediction error of 0.59±0.57 mm, which is on par with the image resolution (0.5mm). Additionally, the standard deviation of our results is lower than the baseline model \cite{bierbrier2023towards}. These signify a robust performance of the FocalErrorNet. One limitation of our work lies in the limited patient data, as public iUS datasets are scarce, while the settings and properties of US scanners can vary, potentially affecting the DL model designs. Therefore, we created random deformations for patch-wise error estimation, and will further explore data-efficient approaches for registration error assessment. 

\section{Conclusion}

We proposed FocalErrorNet, a novel DL model for uncertainty-aware inter-modal registration error estimation in iUS-guided neurosurgery, leveraging the latest focal modulation technique and MC dropout. With thorough assessments of the accuracy and uncertainty measures, we have confirmed the performance of the proposed method against a baseline model previously adopted for the same task. As the first to introduce uncertainty measures and 3D focal modulation in registration error evaluation, our work provides the first step for fast and reliable feedback in inter-modal medical image registration to guide clinical decisions in surgery. We plan to adapt the presented framework for other inter-modal/contrast image registration applications in the future.
\\ \\
\noindent \textbf{Acknowledgment}. We acknowledge the support of the Natural Sciences and Engineering Research Council of Canada (NSERC) and Fonds de Recherche du Québec Nature et Technologies (FRQNT).
 
\bibliographystyle{IEEEtran}
\bibliography{ref.bib}

\end{document}